\documentclass[aps,prd,preprint,a4paper,showpacs,nofootinbib,superscriptaddress]{revtex4-2}
\usepackage[latin9]{inputenc}
\usepackage{babel}
\usepackage{amsmath}
\usepackage{amssymb}
\usepackage{graphicx}
\usepackage[title]{appendix}
\usepackage[unicode=true,
bookmarks=false,
 breaklinks=false,pdfborder={0 0 1},colorlinks=true,linkcolor=red,citecolor=blue]
 {hyperref}
\makeatletter
\@ifundefined{textcolor}{}
{%
 \definecolor{BLACK}{gray}{0}
 \definecolor{WHITE}{gray}{1} 
 \definecolor{RED}{rgb}{1,0,0}
 \definecolor{GREEN}{rgb}{0,1,0}
 \definecolor{BLUE}{rgb}{0,0,1}
 \definecolor{CYAN}{cmyk}{1,0,0,0}
 \definecolor{MAGENTA}{cmyk}{0,1,0,0}
 \definecolor{YELLOW}{cmyk}{0,0,1,0}
}

\usepackage{amsthm}
\usepackage{amsfonts}
\usepackage{bbm}
\usepackage{epsfig}
\usepackage{times}
\usepackage{babel}
\usepackage{color}
\usepackage{framed}
\usepackage{changes}
\usepackage{float}
\usepackage{array}
\usepackage{dcolumn}% Align table columns on decimal pointl
\usepackage{multirow}

\usepackage[margin=1in]{geometry} % 设置页边距
\usepackage{xcolor} % 允许使用颜色
\makeatletter

\@ifundefined{textcolor}{}
{%
 \definecolor{BLACK}{gray}{0}
 \definecolor{WHITE}{gray}{1}
 \definecolor{RED}{rgb}{1,0,0}
 \definecolor{GREEN}{rgb}{0,1,0}
 \definecolor{BLUE}{rgb}{0,0,1}
 \definecolor{CYAN}{cmyk}{1,0,0,0}
 \definecolor{MAGENTA}{cmyk}{0,1,0,0}
 \definecolor{YELLOW}{cmyk}{0,0,1,0}
}

\usepackage{amsthm}
\usepackage{amsfonts}
\usepackage{bbm}
\usepackage{epsfig}
\usepackage{times}
\usepackage{babel}
\usepackage{color}
\usepackage{framed}
\usepackage{changes}
\usepackage{float}
\usepackage{array}

\makeatother

\begin{document}
\title{Probing Dark Matter with Gravitational Waves: Spin-Modulated Dephasing from Black Holes in Halos}
%\title{\blue{Spin-Suppressed Dark Matter Signatures in Extreme Mass Ratio Inspirals}}

\author{Guoyang Fu}
\email{fuguoyang@yzu.edu.cn}
\address{\textit{Center for Gravitation and Cosmology, College of Physical
Science and Technology, Yangzhou University, Yangzhou 225009, China}}

\author{Yunqi Liu}
\email{yunqiliu@yzu.edu.cn (corresponding author)}
\address{\textit{Center for Gravitation and Cosmology, College of Physical
Science and Technology, Yangzhou University, Yangzhou 225009, China}}

\author{Jian-Pin Wu}
\email{jianpinwu@yzu.edu.cn}
\address{\textit{Center for Gravitation and Cosmology, College of Physical
Science and Technology, Yangzhou University, Yangzhou 225009, China}}

\author{Bin Wang}
\email{wang\_b@sjtu.edu.cn}
\address{\textit{Center for Gravitation and Cosmology, College of Physical
Science and Technology, Yangzhou University, Yangzhou 225009, China}}
\address{\textit{Shanghai Frontier Science Center for Gravitational Wave Detection, Shanghai Jiao Tong
University, Shanghai 200240, China}}

\author{Rui-Hong Yue}
\email{rhyue@yzu.edu.cn}
\address{\textit{Center for Gravitation and Cosmology, College of Physical
Science and Technology, Yangzhou University, Yangzhou 225009, China}}

\begin{abstract}

We develop a novel analytical framework for constructing axisymmetric black hole spacetimes sourced by dark matter (DM) halos. Applying this to extreme mass ratio inspirals (EMRIs), we find that the DM induces a detectable gravitational-wave dephasing, scaling monotonically with the halo's compactness. Notably, BH spin significantly suppresses this dephasing, indicating that analyses neglecting rotation would overestimate DM signatures. Faithfulness calculations confirm that future space-borne detectors can robustly distinguish such DM environments, establishing EMRIs as a novel probe for galactic DM distributions.

\end{abstract}
\maketitle

\newpage

\section{Introduction}

Astrophysical and cosmological observations, spanning immense scales, robustly indicate that non-baryonic dark matter (DM) constitutes about 85$\%$ of the total matter in the universe\cite{ Navarro:1995iw,Bertone:2004pz,Freese:2008cz,ParticleDataGroup:2022pth}.
Therefore, identifying its nature remains a central challenge in fundamental physics, with profound implications for our understanding of cosmic structure.
Compelling evidence indicates that a supermassive black hole (SMBH) resides at our Galaxy's core\cite{EventHorizonTelescope:2022wkp,EventHorizonTelescope:2022xqj}. 
Within the intense gravitational field surrounding such SMBHs, matter of diverse forms is expected to accumulate, including significant concentrations of dark matter drawn in through accretion processes and the SMBH's powerful gravitational pull.

Extreme mass ratio inspirals (EMRIs) serve as a powerful probe of strong-field gravity and foundational physics, which establishes their critical importance in gravitational-wave astrophysics. Future space-based observatories such as LISA\cite{Danzmann:1997hm,LISA:2017pwj}, Taiji\cite{Hu:2017mde}, and TianQin\cite{TianQin:2015yph,Li:2024rnk} are poised to detect the GW signals from these systems, offering a potent means to test general relativity and characterize the matter environment adjacent to SMBHs.
By precisely encoding the underlying spacetime structure, these signals enable stringent constraints on alternative gravity models and physical conditions near SMBHs.
This local environment exerts ``environmental effects'', modifying the compact object's inspiral dynamics and the emitted GW signature.
These effects collectively produce measurable deviations in the GW waveform profile.
Therefore, EMRI observations, through the decoding of their intricate waveforms, furnish a distinctive opportunity to probe the density profile and nature of dark matter clustering around SMBHs.

Recent work within spherically symmetric gravitational frameworks has shown that a DM halo can imprint measurable phase deviations on EMRI gravitational waveforms \cite{Cardoso:2021wlq,Figueiredo:2023gas,Dai:2023cft,Zhang:2024ugv}.
These deviations arise from altered orbital decay rates induced by dynamical friction and gravitational disturbances caused by the DM halo. 
As a result, the gravitational waveform's phase evolution exhibits sensitivity to both the halo's density and radial scale.
Such imprints become particularly significant in environments with high density of DM near the SMBH. 
This establishes GW observations as an innovative approach for constraining DM parameters.

%%%%%%%%%%%%%%%%%%%%%%%%%%%%%%%%%%%%%%%%%%%%%%%%%%%%%%%%%%%%%%%%%%%%

The prevalent assumption of spherical symmetry in modeling dark matter near black holes contrasts with the typical rotation of astrophysical black holes.
Spin-induced frame-dragging in a Kerr spacetime can generate azimuthal DM flows proportional to the spin parameter\cite{Sadeghian:2013laa}, even when the initial DM configuration is spherically symmetric.
%Prior investigations have revealed key dynamics: Ref.{XXX} examined geodesic motion of dark matter around established black holes, identifying distinctive signatures in annihilation spectra. 
%\xg{In \cite{Ferrer:2017xwm} the authors studied the impact of black hole spin on the density profile of DM in Kerr spacetime}{This rotation significantly alters the DM density profile}, \xg{quantifying how BH rotation reshapes the surrounding DM spike.}{effectively reshaping the surrounding DM spike\cite{Ferrer:2017xwm}.} 
This rotation significantly alters the DM density profile, effectively reshaping the surrounding DM spike\cite{Ferrer:2017xwm}.
%Furthermore, recent investigations indicates that DM can have a suppressive influence on black hole superradiance \cite{Liu:2024qso}.
Further investigations of rotating black holes within dark matter environments appear in \cite{Liu:2024qso}.
Motivated by these findings, we develop a novel analytical formalism for generating axisymmetric black hole solutions immersed in dark matter halos, allowing for a study of DM imprints through EMRI signals.
Our analysis identifies characteristic imprints of DM environment in gravitational waveforms, with the goal of systematically contrasting the signals produced by two distinct density profiles.

% 1EMRI 2 evironmental effect. 3 sphericallg symmetric 4 letter exetension

\section{Theoretical setup}
In this section, the geometry induced by the supermassive black hole surrounded by DM halo will be constructed.
%We will assume a symmetry of the spacetime, then use some conditions to constraint the matter distribution, and finally get a  solution to the spacetime.
Adopting a ``Kerr-like" symmetric ansatz, we describe the geometry through the line element:
\begin{eqnarray}\label{metricfun}
ds^2&=&\big[-1+\frac{f(r)}{\Sigma}\big]dt^2+\frac{\Sigma}{\Delta}dr^2-\frac{2a f(r)\sin^2\theta}{\Sigma}dtd\phi\nonumber\\&+&\Sigma d\theta^2
+\sin^2\theta\big[a^2+r^2+\frac{a^2f(r)\sin^2\theta}{\Sigma}\big]d\phi^2,
\end{eqnarray}
where $a$ denotes the rotation of the black hole, $\Sigma=r^2+a^2 \cos\theta^2$, and $\Delta=r^2+a^2-B(r)$. The undetermined functions $f$ and $B$ depend only on the radial coordinate $r$.

By Einstein equation $G_{\mu\nu}=T_{\mu\nu}$, we could write out the energy-momentum tensor $T_{\mu\nu}$ in terms of metric functions in \eqref{metricfun}. 
To analyze the properties of the DM, we project the energy-momentum tensor onto the co-moving observer's orthonormal tetrad frame which behaves as
\begin{eqnarray}
e^{\mu}_{(0)}&=&\Big(\frac{r^2+a^2}{\sqrt{\Sigma(r^2+a^2-f(r))}},0,0,\frac{a}{\sqrt{\Sigma(r^2+a^2-f(r))}}\Big),\nonumber\\
e^{\mu}_{(1)}&=&\Big(0,\frac{r^2+a^2-f(r)}{\sqrt{\Sigma}},0,0\Big),\nonumber\\
e^{\mu}_{(2)}&=&\Big(0,0,\frac{1}{\sqrt{\Sigma}},0 \Big),\nonumber\\
e^{\mu}_{(3)}&=&\Big(-\frac{a\sin\theta}{\sqrt{\Sigma}},0,0,-\frac{1}{\sqrt{\Sigma}\sin\theta}\Big)\nonumber.
\end{eqnarray}
\
\\

%\subsection{$p_r$}
The projection of $T_{\mu\nu}$ onto the space-like tetrad $e^{\mu}_{(1)}$ gives the pressure along the radial direction,
\begin{eqnarray}\label{}
p_r=G_{\mu\nu}e^{\mu}_{(1)}e_{\nu}^{(1)}
     =\frac{1}{X_1}(H_1+a^2 K_1+a^4 P_1)\nonumber
\end{eqnarray}
with
\begin{eqnarray}\label{co-pr}
X_1&=&\Sigma^3 (r^2-f(r)+a^2),\nonumber\\
H_1&=&r^2B(r)(f(r)+r(r-f'(r)))+r^4(r f'(r)-2f(r)),\nonumber\\
K_1&=&(2+2\cos2\theta)f(r)r^2-\cos\theta^2f(r)^2\nonumber\\&-&r[B(r)\cos\theta^2(r-f'(r))+(1+\cos\theta^2)r^2f'(r)],\nonumber\\
P_1&=&\cos\theta^2(f(r)-r f'(r)). \nonumber
\end{eqnarray}

In the non-rotating limit, the spacetime geometry reduces to spherical symmetry, and our model recovers a Schwarzwald BH harbored in a DM halo.
For cases with spherical symmetry, for example, in \cite{Cardoso:2021wlq,Figueiredo:2023gas,Dai:2023cft, Zhao:2024bpp}, the authors followed an ``Einstein cluster'' \cite{Einstein:1939ms,Geralico:2012jt} and generalized it to include a central Schwarzwald BH.
This figuration assumes an anisotropic matter distribution with only a tangential pressure and vanishing radial pressure.
In this work, we also employ an Einstein cluster to enforce the condition of vanishing radial pressure in the limit of zero black hole spin ( $a \to 0$ ), i.e., $\lim_{a\rightarrow 0} p_r=0$.
This condition leads to
\begin{eqnarray}\label{fder}
f'(r)=\frac{2f(r)r^2-B(r)(f(r)+r^2)}{r^3-r B(r)}.
\end{eqnarray}
Rewriting $B(r)=2r m(r)$, one gets
\begin{eqnarray}\label{B}
m(r)=\frac{2rf(r)-r^2f'(r)}{2(r^2+f(r)-rf'(r))}.
\end{eqnarray}
Actually, $m(r)$ describes the mass distribution.
Once we have an expression for the mass distribution function $m(r)$, by solving \eqref{B} we get the function $f(r)$.
Then substituting functions $f(r)$ and $m(r)$ into metric \eqref{metricfun}, we could get the solutions of the Kerr-like spacetime \eqref{metricfun}.

The energy-momentum tensor $T^{\mu}_{\nu}$ in the spacetime can be expressed as 
\[
\left ( \begin{array}{cccc}
-\rho & 0 & 0 & p_4 \\
0 & p_r & 0 & 0 \\
0 & 0 & p_t & 0 \\
p_4 & 0 & 0 & p_t
\end{array}\right )
\]
The non-vanishing components could be obtained by projecting the Einstein tensor onto the comoving observer's orthonormal tetrad frame. 
As a result, the functions $\rho$, $p_t$, and $p_4$ can be expressed as complicated functions of $f(r)$ and $B(r)$.
In the non-rotating limit as $a\rightarrow0$, one reaches $\rho_0(r)=\rho(r)_{a\rightarrow0}=\frac{2m'(r)}{r^2}$.
Thus, from the spherically symmetric distribution of the mass density $\rho_0(r)$ we could solve $Y=\exp({\int{Q dr}})$ with $Q^{-1}=\int{(1-\rho_0 r^2)}dr$ and $m(r)=(r-Y)/2$.
During the integral, the boundary condition takes as $m(r_h)=m_{BH}$ where $m_{BH}$ is the mass of the central BH.
Furthermore, together with \eqref{fder} and \eqref{B}, we can get the function $f(r)$ and $B(r)$.
Finally we obtain the information of the distribution of the dark matter and the metric in the Kerr-like spacetime.

In the following sections, we will consider certain type of spherically dark matter distribution profile $\rho_0(r)$ to solve the spacetime, after that we will investigate the EMRI with such rotating supermassive BH in the DM environment.
%\subsection{Hernquist-type profile}
%We consider a Hernquist-type density distribution $\rho_0=\frac{M a_0}{2\pi r (r+a_0)^3}$ where $M$ is the total mass of the ``halo" and $a_0$ a typical lengthscale. Other popular density profiles Navarro-Frenk-White, Jaffe or King profiles \cite{XXX-cardoso}.
%Follow the cardoso's work,  the mass distribution as 
%\begin{eqnarray}\label{eq-mass}
%m(r)=m_{BH}+\frac{M r^2}{(a_0+r)^2}(1-\frac{2m_{BH}}{r})^2.
%\end{eqnarray}
%With Eq.\eqref{eq-mass}, one could fix the metric function $f(r)$ and $B(r)$.
%Setting $a_0=10^4 m_{BH}$, $M=100m_{BH}$, we plot the metric functions in Fig.\ref{}.

\section{Dark matter density profiles}
We consider a parameterized density profile for galactic dark matter distributions, governed by parameters $(\alpha, \beta, \gamma)$ \cite{Graham:2005xx,Taylor:2002zd}, which takes the form as
\begin{equation}\label{DMprofile}
\rho_0(r)=\rho_c(r/a_0)^{-\gamma}[1+(r/a_0)^\alpha]^{(\gamma-\beta)/\alpha}.
\end{equation}
Here, $a_0$ defines the characteristic scale radius of the dark matter halo and the $\rho_c$ denotes the central density. 
The parameters $\beta$ and $\gamma$ govern the slopes at large ($r\gg a_0$) and small ($r\ll a_0$) distances, respectively, while $\alpha$ determines the sharpness of the transition. 
%For $(\alpha, \beta, \gamma)=(1,4,1)$ and $(\alpha, \beta, \gamma)=(1,3,1)$, the density profile \eqref{DMprofile} corresponds to the Hernquist (HQ) profile \cite{Hernquist:1990be} and Navarro-Frenk-White (NFW) profile \cite{Navarro:1996gj}, respectively. 
For $(\alpha, \beta, \gamma)=(1,4,1)$ the density profile \eqref{DMprofile} corresponds to the Hernquist (HQ) profile \cite{Hernquist:1990be}, while $(\alpha, \beta, \gamma)=(1,3,1)$ the Navarro-Frenk-White (NFW) profile \cite{Navarro:1996gj}. 
In the NFW model, to avoid the logarithmic divergence of the mass function $m(r)$ we introduce a radial cut-off $\rho_0(r>r_c)=0$ with $r_c=5a_0$. 

%%%%%%%%%%
\begin{figure}[h!]
	\center{
	\includegraphics[scale=0.46]{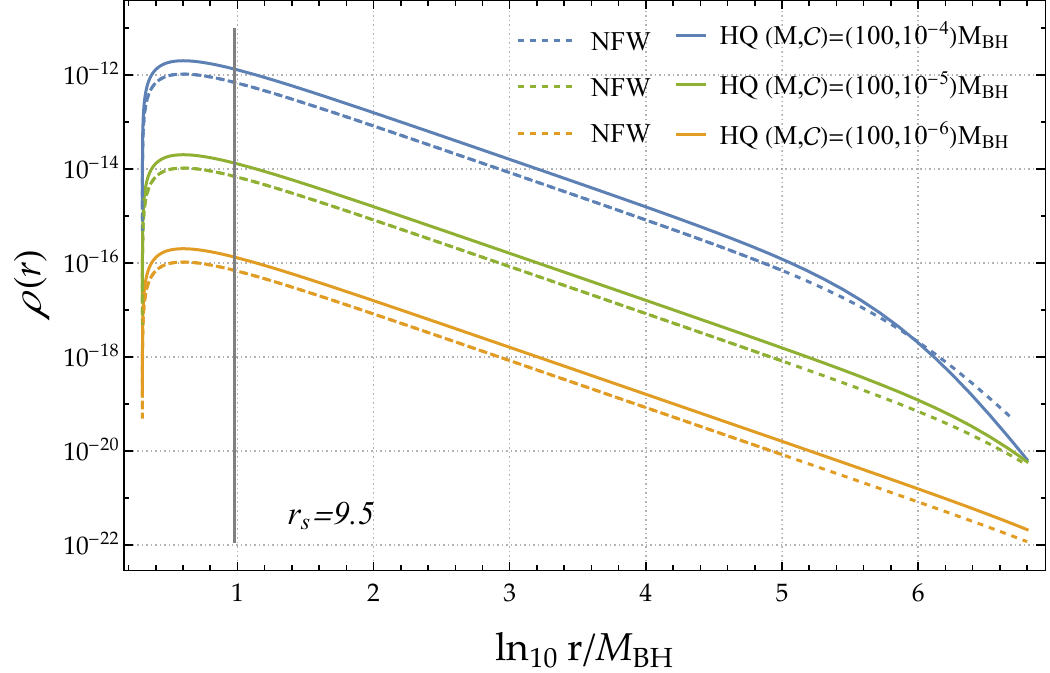}
        \caption{The DM density distributions $\rho(r)$ as a function of the radial $r$ for the HQ and NFW models with different configurations. The gray vertical line represents the initial orbital separation $r_s$ between the central supermassive black hole and the secondary object.}
		\label{DMprofile}
	}
\end{figure}
%%%%%%%%%%%

The relativistic DM solution, derived from the vacuum solution, depends only on the two parameters of the mass of the DM $M$ and the compactness $\mathcal{C}=M/a_0$. 
In Milky Way-like galaxies, the surrounding DM halos are predicted to possess masses substantially exceeding those of their central supermassive black holes (i.e. $M \gg m_{BH}$), while exhibiting remarkably low compactness parameters typically characterized by $\mathcal{C}\leq 10^{-4}$ \cite{Navarro:1996gj}.
Therefore, we fixed the DM mass as $M=100 m_{BH}$ and investigated the observable effects of low compactness under the axisymmetric spacetime \eqref{metricfun}.
Fig. \ref{DMprofile} displays the dark matter halo density distributions $\rho_0(r)$ for different profiles.

\section{Trajectory and Waveform}
In this section, we investigate the trajectory of the secondary compact object and establish the Kludge waveform using the \texttt{FastEMRIWaveforms} (FEW) package \cite{Katz:2021yft,Chua:2020stf}. 

First, we regard the secondary compact object as a point particle and restrict its orbits to lie in the equatorial plane ($\theta=\pi/2$). In this situation, the conserved energy $E$ and the angular momentum $L_z$ of the system can be expressed as
\begin{eqnarray}
\label{energy}
E&=&-u_t=-g_{tt}u^t-g_{t\phi}u^{\phi},  \\
\label{momentum}
L_z&=&u_{\phi}=g_{t\phi}u^t+g_{\phi\phi}u^\phi,
\end{eqnarray}
where $u^{\mu}$ is the 4-velocity. For the bounded orbits, we also introduce the eccentricity $e$ and semilatus $p$ to parameterize the conserved quantities, which are defined as:
\begin{eqnarray}
e=\frac{r_a-r_p}{r_a+r_p}, \ p=\frac{r_ar_p}{r_a+r_p},
\end{eqnarray}
where $r_a$ and $r_p$ correspond to the apastron and periastron, respectively.  
Using the 4-velocity condition $g_{\mu\nu}u^{\mu}u^{\nu}=-1$, the Eqs.\eqref{energy} and \eqref{momentum} can be solved numerically. To efficiently interpolate, we 
follow the method in \cite{Katz:2021yft,Chua:2020stf} and define a new parameter $u=\ln{(p-p_s+3.9)}$ to build a uniform grid $(u,e)$, where $p_s$ is the separatrix for the Kerr spacetime. This approach allow us to explore more points closer to the separatrix. In the absence of dark matter, the orbital frequencies reduce to that of a timelike geodesic in vacuum Kerr spacetime, in agreement with the \texttt{KERRGEODESICS} results \cite{Fujita:2009bp}, thereby verifying the consistency of our model.

%Combining with the eqs. \eqref{energy} and \eqref{momentum}, the evolution of the inspiral trajectory can be computed by the numerically integrating. 

Gravitational radiation carries away orbital energy and angular momentum, thereby influencing the trajectories of point particle. To calculate the energy and angular fluxes, we consider adiabatic approximation \cite{Ryan:1995zm,Babak:2006uv} and inerpolate the fluxes with bicubic splines on the grid $(u,e)$. 
In the weak field approximation, the gravitational radiation is given by the quadrupole formula
\begin{eqnarray}
\left<\dot{E}\right>&=&-\frac{1}{5}\left<I_{TT}^{jk(3)}I_{TT}^{jk(3)}+\frac{16}{9}J_{TT}^{jk(3)}J_{TT}^{jk(3)}\right>, \nonumber\\
\left<\dot{L}_z\right>&=&-\frac{2}{5}\epsilon^{3kl}\left<I_{TT}^{ka(3)}I_{TT}^{al(3)}+\frac{16}{9}J_{TT}^{ka(3)}J_{TT}^{al(3)}\right>, \nonumber
\end{eqnarray}
where the superscript number denotes the order of the time derivative. The mass moment $I^{jk}$ and the angular momentum moments $J^{jk}$ are
\begin{eqnarray}
I^{jk}&=&\mu x^j x^k, \\
J^{jk}&=&\epsilon_{jml} v^m I^{lk}.
\end{eqnarray}
Here, the Cartesian coordinates $x^i$ can be defined as $x^i=\{r \cos(\phi),r\sin(\phi),0\}$.

We numerically integrate the energy and angular momentum fluxes over the quasi-Keplerian true anomaly $\chi$ from $0$ to $2\pi$ and interpolate the fluxes with bicubic splines on the grid $(u, e)$. It is noteworthy that, to avoid error magnification in the interpolation, we interpolate over the effective fluxes residual after subtract out the lowest order post-Newtonian contribution on the $(u, e)$ plane. 

%By fixing the initial eccentricity $e_0=0.1$, $m_{BH}=10^6M_\odot$, $m_p=10M_\odot$ and adjusting the initial semi-latus rectum $p_0$ to ensure the evolution period remains one year before the plunge at $p_{end}=p_s+0.1$. From Fig. \ref{pt}, we see that, when we introduce the rotation parameter $a$, the $p_0$ become smaller compared to the case of $a=0$. This implies that the EMRI embedded in rotating DM-BH exhibits slower orbital decay. For more detail, we list the values of the initial $p_0$ in Table \ref{tab_p}. Furthermore, we observed that the parameter $a$ does not affect the relational behavior between the different DM profiles and configurations, indicating that the parameter $a$ independence of the DM parameters $M$ and $a_0$. 

Now, we turn to investigate the number of orbital cycles $\Delta \mathcal{N}$ for EMRIs embedding in the DM halo. For a given observational time $t_f$, we can compute the the number of orbital cycles $\Delta \mathcal{N}$ by integrate the orbital frequencies as \cite{Maselli:2020zgv,Barsanti:2022ana}
\begin{eqnarray}
\Delta\mathcal{N}=\int_{t_i}^{t_f} \Delta\dot{\phi}(t) dt
\end{eqnarray}
where  $\Delta\dot{\phi}(t)$ denotes the difference in orbital frequency between the cases with and without a DM halo,  defined as $\Delta\dot{\phi}_i(t)=|\dot\phi^{DM}-\dot\phi^{Vacuum}|$. 
We adopt the LISA detection threshold $\Delta \mathcal{N} \sim 1$ rad as the criterion for dephasing \cite{Gupta:2021cno}.

\begin{figure}[H]
	\center{
	\includegraphics[scale=0.43]{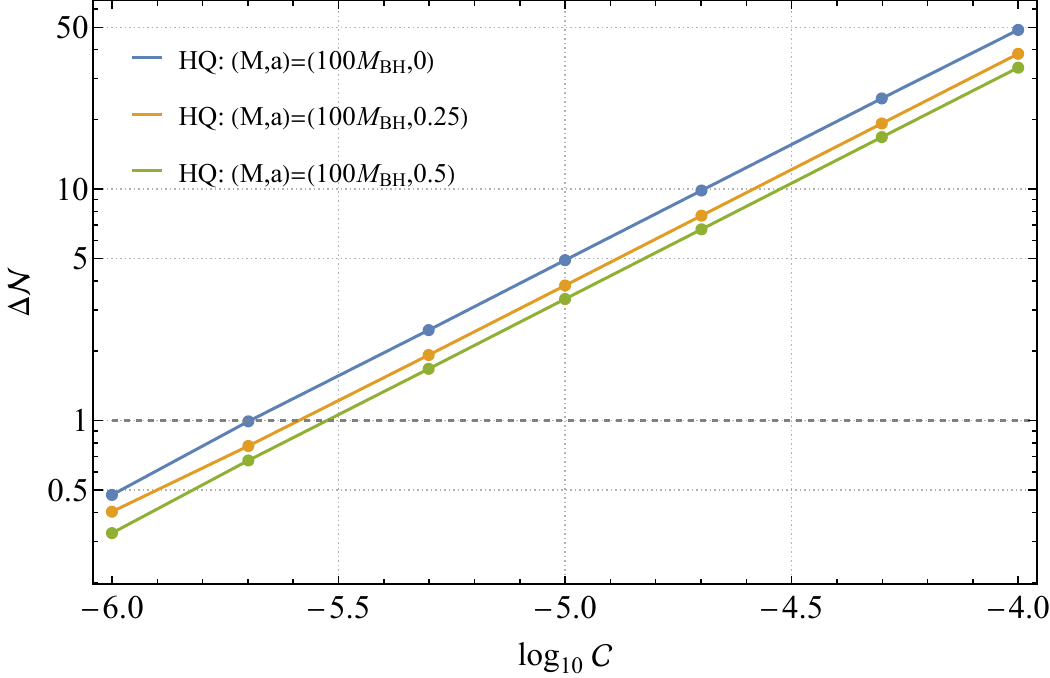}		
        \caption{The difference in the number of orbital cycles, $\Delta \mathcal{N}$, with versus without a DM halo, plotted as a function of halo compactness $\mathcal{C}$ for both non-rotating and rotating BHs. The Dashed line represents the threshold  $\Delta \mathcal{N}=1$ rad.}
		\label{fig_2}
	}
\end{figure}

For the fixed system parameters $e_0=0.1$, $m_{BH}=10^6M_\odot$ and $m_p=10M_\odot$, the number of orbital cycles $\Delta \mathcal{N}$ is numerically evaluated as demonstrated in Fig. \ref{fig_2}. 
Over a one-year observation, $\Delta \mathcal{N}$ increases monotonically as the halo compactness $\mathcal{C}$ decreases. This indicates that a denser, more compact DM halo produces a stronger cumulative phase shift, enhancing its detectability. 
However, introducing black hole spin ($a>0$) leads to a significant suppression in $\Delta \mathcal{N}$ , demonstrating that neglecting the rotation effects would systematically overestimate the GW signal.     
%%%%%%%%%%
%%%%%%%%%%%

Furthermore, we employ the faithfulness $\mathcal{F}$ to assess environmental effects from different DM distributions.
For any pair of waveforms, it is defined as
\begin{eqnarray}\label{faithfulness}
\mathcal{F}\left[h_{v}, h_{m}\right]=\max _{\left\{t_{c}, \phi_{c}\right\}} \frac{\left\langle h_{v} \mid h_{m}\right\rangle}{\sqrt{\left\langle h_{v} \mid h_{v}\right\rangle\left\langle h_{m} \mid h_{m}\right\rangle}}\,,
\end{eqnarray}
where the subscripts $v$ and $m$ generically label the waveforms being compared. This allows us to quantify differences both between any DM halo and vacuum, and between different DM halos themselves.
The noise-weighted inner product is defined as
\begin{eqnarray}
\left\langle h_{m} \mid h_{v}\right\rangle & = & 4 \Re\int_{f_{\min }}^{\max } \frac{\tilde{h}_{m}(f) \tilde{h}_{v}^{*}(f)}{S_{n}(f)} d f  ,
\end{eqnarray}
where $\tilde{h}(f)$ is the Fourier transform of $h(t)$, the asterisk denotes complex conjugation, and $S_n(f)$ is the noise power spectral density of LISA \cite{Robson:2018ifk}. 
Following the distinguishability criterion \cite{Lindblom:2008cm,Flanagan:1997kp,Chatziioannou:2017tdw}, two waveforms are considered distinguishable if the faithfulness falls below $1-d/(2\text{SNR}^2)$, where $d$ is the parameter dimension of the EMRI system. 

\begin{figure}[H]
	\center{
	\includegraphics[scale=0.43]{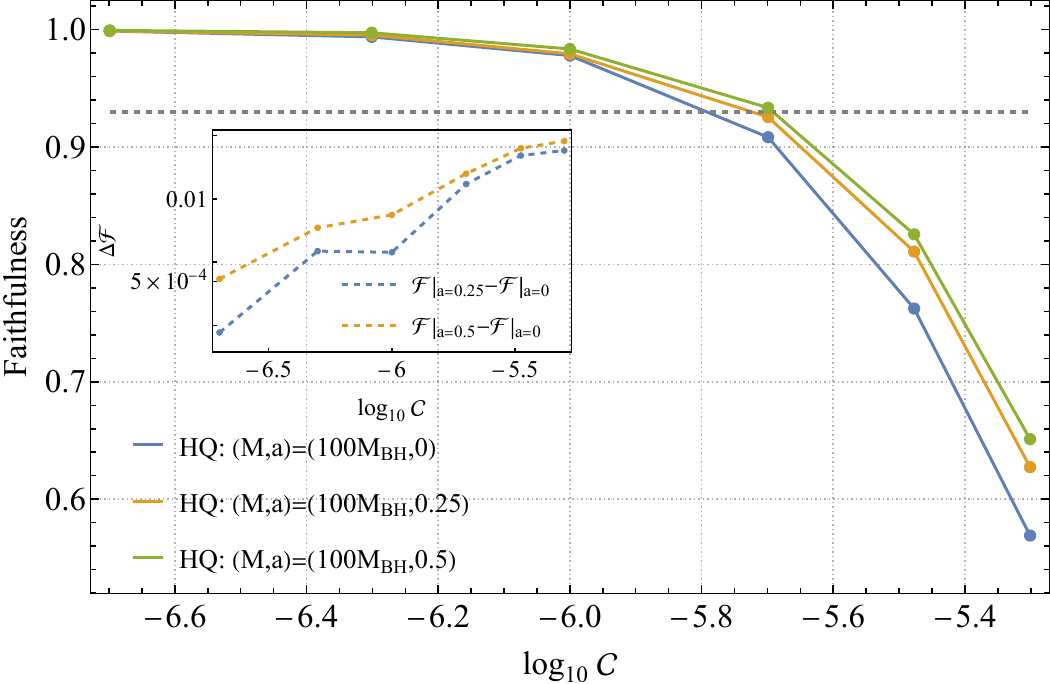}		
        \caption{The faithfulness between waveforms with and without a DM halo, after one-year accumulation for different parameters $a$ and $\mathcal{C}$. The dashed line represents the distinguishability threshold  $1-d/(2\text{SNR}^2) \simeq  0.92$.}
		\label{fig_3}
	}
\end{figure}

Fig.\ref{fig_3} depicts the faithfulness of the Hernquist-type DM halo against vacuum, demonstrating a monotonic decrease with $\mathcal{C}$.
A faithfulness threshold indicates that halos with $\mathcal{C} > 10^{-5}$ yield distinguishable EMRI waveforms. Importantly, our analysis of the faithfulness further confirms the significant suppressive effect of BH spin on DM signatures noted previously.

%%%%%%%%%%
%%%%%%%%%%%

	%%%%%%%%%%%%%%%%%%%%%%%%%%%%%%%%%%%%%
\begin{table}[H]
	\centering
	\begin{tabular}{|c|c|c|c|c|}
		\hline
		$a$  &  $\mathcal{C}$  & HQ-Vacuum  &  NFW-Vacuum  & HQ-NFW  \\
		\hline
		\multirow{3}{*}{0} &  $10^{-5}$  &  0.212  &  0.246  &  0.954  \\
		\cline{2-5}
		&   $5\times 10^{-5}$   & 0.908   &  0.896  &    0.990    \\
		\hline
		\multirow{3}{*}{0.25} &   $10^{-5}$   &  0.283  & 0.307  &    0.976     \\
		\cline{2-5}
		&   $5\times 10^{-5}$   & 0.925   &  0.903  &    0.986    \\
		\hline
	\end{tabular}
	\caption{Faithfulness for different DM profiles compared to vacuum (HQ-Vacuum, NFW-Vacuum) and compared to each other (HQ-NFW), with parameters $a = 0, 0.25$ and $\mathcal{C} = 10^{-5},\,5\times10^{-5}$.  }	\label{faithfulness_tab}
\end{table}
%%%%%%%%%%%%%%%%%%%%%%%%%%%%%%%%%%%%%%%

Extending the analysis to different DM density profiles, we find that for the $\mathcal{C} = 10^{-5}$ (see Table.~\ref{faithfulness_tab}), the faithfulness of the HQ-Vacuum and NFW-Vacuum is significantly lower than the detection threshold of $\simeq 0.92$, indicating that LISA can easily distinguish DM environments from vacuum. However, for the HQ-NFW case, the faithfulness increases significantly compared to the HQ-Vacuum and NFW-Vacuum cases, exceeding the detection threshold. This suggests that more compact density profiles may be required to distinguish different DM environments. Our results demonstrate that the LISA is less sensitive to distinguishing between different DM environments, but more sensitive to the presence of DM effects.

\section{Summary and discussion}
This study investigates the imprint of DM on gravitational waves from EMRIs around rotating supermassive black holes. We first construct Kerr-like black hole solutions surrounded by DM halos, parameterized by HQ and NFW density profiles.
By enforcing the Einstein cluster condition (vanishing radial pressure as $a\rightarrow0$), we solve the metric functions.
Using the FEW package, we simulate EMRI trajectories and kludge waveforms for equatorial orbits.
The orbital decay and gravitational radiation fluxes are computed via adiabatic approximation, with initial conditions adjusted for a one-year evolution before plunge.
Our results establish EMRIs as powerful tools to probe DM environments around rotating black holes.
The dephasing $\Delta \mathcal{N}$ and faithfulness $\Delta \mathcal{F}$ provide observable metrics sensitive to DM density profiles and compactness. Crucially, black hole spin diminishes DM-induced phase shifts, necessitating self-consistent rotating spacetime models for accurate parameter estimation. 
Future space-based detectors will be capable of constraining DM halos with $\mathcal{C}\geq10^{-5}$, offering insights into DM accretion physics and halo morphology. Limitations include the neglect of dynamical friction beyond adiabatic approximations; future work should incorporate self-consistent radiation reaction and non-equatorial orbits.
 
{\it \textbf{Acknowledgments.}}
This work is supported by National Key R$\&$D Program of China (Nos. 2023YFC2206703 and 2020YFC2201400), the Natural Science Foundation of China under Grants Nos. 12347159, 12375055 and 12375056.
B. W. was partially supported by NNSFC under grant 12075202.

\begin{appendices}

\end{appendices}

 \end{document}